\documentclass[twocolumn,showpacs,preprintnumbers,amsmath,amssymb,prb]{revtex4}

\usepackage{graphicx}
\usepackage{dcolumn}
\usepackage{bm}
\usepackage{amsmath}

\begin{document}


\title{Transport gap in side-gated graphene constrictions}

\author{F. Molitor}
\email{fmolitor@phys.ethz.ch}
\author{A. Jacobsen}
\author{C. Stampfer}
\author{J. G\"uttinger}
\author{T. Ihn}
\author{K. Ensslin}
\affiliation{Solid State Physics Laboratory - ETH Zurich, Switzerland}

\date{\today}

\begin{abstract}

We present measurements on side gated graphene constrictions of different geometries. We characterize the transport gap by its width in back gate voltage and compare this to an analysis based on Coulomb blockade measurements of localized states. We study the effect of an applied side gate voltage on the transport gap and show that high side gate voltages lift the suppression of the conductance. Finally we study the effect of an applied magnetic field and demonstrate the presence of edge states in the constriction.

\end{abstract}

\maketitle



\section{Introduction}

Graphene has unique electronic properties due to its special band structure and Dirac-like behavior of the electronic quasi-particles.\cite{novoselov_nature2005, zhang_nature2005} However, this specific structure has one main draw-back: because of the absence of a band  gap and the prediction of Klein tunneling,\cite{katsnelson_naturephys2006} it is difficult to confine carriers electrostatically. Recently this problem has been overcome by cutting a graphene flake into narrow ribbons, which has been shown to open a gate voltage region of suppressed conductance, called transport gap.\cite{chen_physE2007,han_prl2007} This method can be used to confine carriers in quantum dots.\cite{ponomarenko_science2008, stampfer_apl2008, schnez_condmat2008} However, even if by now graphene constrictions are regularly used in combination with quantum dots, either to form the tunnel barriers or as a charge readout detector,\cite{guttinger_condmat2008} the origin of the formation of the transport gap is still not well understood. The most common picture is that transport is suppressed due to edge disorder leading to localization.\cite{sols_prl2007,mucciolo_condmat2008,evaldsson_condmat2008} One possibilty to study the influence of the edges experimentally is to use graphene side gates, as they mainly act on the edges of the structure. Graphene side gates have already proven to be useful in combination with many different graphene structures, for example to study the homogeneity in a graphene Hall bar,\cite{molitor_prb2007} to individually tune the tunnel barriers and the dot energies in a graphene quantum dot,\cite{stampfer_nanolett2007} and to locally control the arms of an Aharonov-Bohm ring.\cite{molitor_toappear}

In this work, we study the dependence of the transport gap on the constriction geometry and the influence of an applied side gate voltage. Finally we analyze the effect of an external magnetic field on transport through the constriction.
 

\section{Sample and setup}

\begin{figure}
\includegraphics[width=0.45\textwidth]{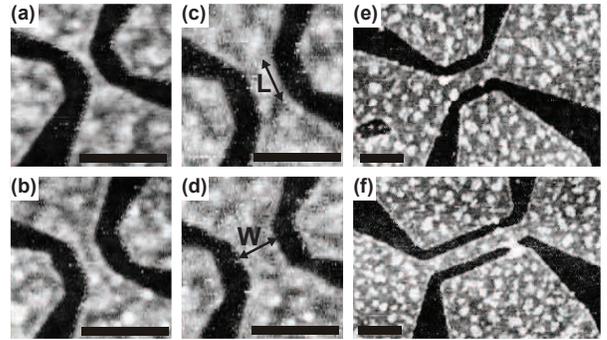}
\caption{\label{fig:Fig1} Scanning force micrographs (SFM) of the constrictions studied in this work. The length of the scale bars is 200 nm. The dimensions are the following: (a) $W$=30~nm, $L$=100~nm (b) $W$=65~nm, $L$=100~nm (c) $W$=75~nm, $L$=100~nm (d) $W$=100~nm, $L$=100~nm (e) $W$=85~nm, $L$=200~nm (f) $W$=85~nm, $L$=500~nm. }
\end{figure}

Fig. \ref{fig:Fig1} displays scanning force micrographs of the graphene constrictions studied in this work. The graphene flakes are produced by mechanical exfoliation of graphite and deposition on a highly doped Si substrate, covered by 295~nm of silicon dioxide.\cite{novoselov_science2004} Raman spectroscopy is used in order to confirm the single-layer nature of the flakes.\cite{ferrari_prl2006,graf_nanolett2007} The structure is defined by electron beam lithography followed by reactive ion etching. Finally, contacts are added using a second electron beam lithography step and Cr/Au evaporation. Each structure consists of source and drain leads, connected by the constriction, and has two lateral side gates located $\approx50$~nm from the constriction. Structures (a) to (d) from Fig. \ref{fig:Fig1} have the same constriction length, but different widths, and are etched out of the same graphene flake. Structures (e) and (f) have the same width and different lengths and originate from a second flake on the same substrate.

The measurements are performed in a variable temperature inset cryostat at $T\approx 2$~K. For constant bias measurements, a dc bias voltage $V_{\mathrm{bias}}=300~\mu$V is applied symmetrically across the structure, and the current is measured. In the case of measurements as a function of $V_{\mathrm{bias}}$ and $V_{\mathrm{BG}}$, the conductance is measured directly by low-frequency lock-in techniques by adding an ac bias of 100~$\mu$V to the dc contribution. 


\section{Results and discussion}


\subsection{Characterization of the transport gap}\label{sec:tg}

\begin{figure}
\includegraphics[width=0.45\textwidth]{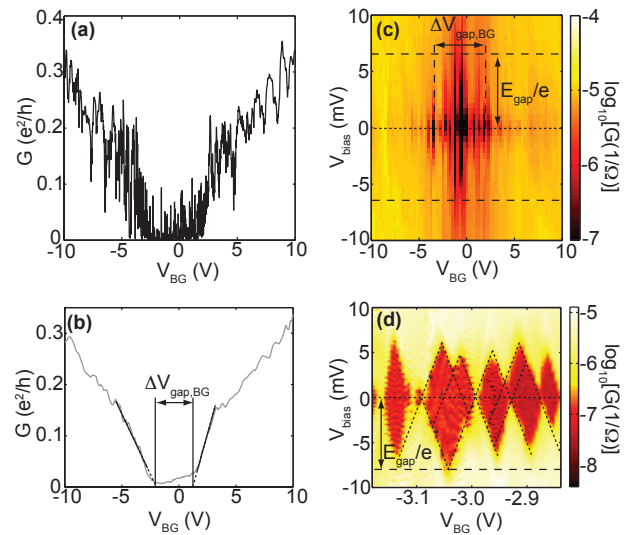}
\caption{\label{fig:Fig2} Transport gap measurements for the constriction with dimensions $W$=85~nm, $L$=500~nm (a) Conductance versus back gate voltage $V_{\mathrm{BG}}$. The measurement is performed with a bias voltage of $300~\mu$V. (b) Procedure used to determine the size of the transport gap in back gate voltage $\Delta V_{\mathrm{gap,BG}}$. The trace from (a) is smoothened over 2.5 V in $V_{\mathrm{BG}}$. The black lines indicate the linear fits used to determine the gap size. (c) Color plot of the conductance as a function of applied back gate and bias voltage. (d) Zoom of the gap measured in (c).}
\end{figure}

Fig. \ref{fig:Fig2}(a) shows the conductance through the $W$=85~nm, $L$=500~nm constriction as a function of applied back gate voltage. Around $V_{\mathrm{BG}}\approx0$~V, there is a region of strongly suppressed conductance, the so-called transport gap. However, it is not a real energy gap because it is full of sharp resonances. The  size of this transport gap in back gate voltage can be quantified by the procedure shown in Fig. \ref{fig:Fig2}(b). The conductance trace is smoothened over a back gate voltage range large enough to eliminate the resonances without affecting the general shape. The regions of a linear increase of the conductance at both sides of the transport gap are selected by hand, and a linear fit is performed (black lines). The gap size in back gate voltage $\Delta V_{\mathrm{gap,BG}}$ is then defined as the distance between the intersection points of the fitted traces with the $G=0$ line ($\Delta V_{\mathrm{gap,BG}}=3.4$~V). This is a reasonable approach since the conductance values are much smaller than the minimal conductivity observed for extended graphene systems which is of the order of $4e^2/h$.\cite{novoselov_nature2005} We also tried different approaches to define the gap, as for example by defining a cutoff current. The overall results were the same, even if the details changed slightly.

Fig. \ref{fig:Fig2}(c) displays the conductance as a function of $V_{\mathrm{bias}}$ and $V_{\mathrm{BG}}$. A region of suppressed conductance can again be observed, with an extension $E_{gap}/e$ in $V_{\mathrm{bias}}$-direction and $\Delta V_{\mathrm{gap,BG}}$ along the direction of $V_{\mathrm{BG}}$. 
By recording a higher resolution measurement of the region where the transport gap is most pronounced (Fig. \ref{fig:Fig2}(d)), one can see that the region of supressed conductance is composed of individual diamonds, sometimes overlapping as in the case of statistical Coulomb blockade.\cite{dorn_prb2004} The gap position is shifted in back gate voltage compared to Fig. \ref{fig:Fig2}(a),(b),(c) due to a change of the sample between the corresponding measurements. The energy $E_{\mathrm{gap}}$ corresponds to the charging energy of the largest diamond. By using the back gate lever arm $\alpha_{\mathrm{BG}}\approx0.084$ determined from the diamond measurements, one gets two different energy scales that differ by more than an order of magnitude ($E_{\mathrm{gap}} \approx 8$~meV, $\alpha \Delta V _{\mathrm{gap,BG}}\approx285$~meV). The gap $E_{\mathrm{gap}}$ is a measure for the charging energy of the smallest electron puddle that controls the transmission. The quantity $\alpha \Delta V _{\mathrm{gap,BG}}$ is a measure of the energy interval in which localized electrons govern transport at the Fermi energy.\cite{stampfer_inprep} The physics is governed by statistical Coulomb blockade, but the number of puddles is small enough to prevent complete self-averaging.

\subsection{Influence of the constriction geometry}\label{sec:cg}

\begin{figure}
\includegraphics[width=0.45\textwidth]{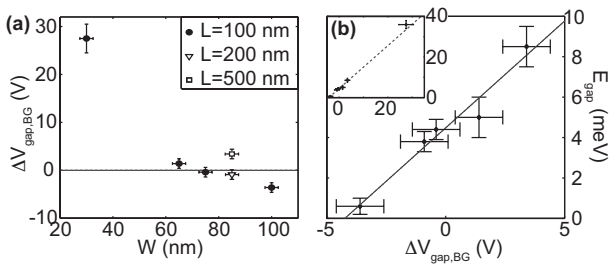}
\caption{\label{fig:Fig3} (a) $\Delta V_{\mathrm{gap,BG}}$, determined by the procedure described in Fig. \ref{fig:Fig2}(b), as a function of the constriction width. The dashed line indicates $\Delta V_{\mathrm{gap,BG}}=0$~V. (b) $E_{\mathrm{gap}}$, determined by taking the energy of the largest diamond as shown in Fig. \ref{fig:Fig2}(d), as a function of $\Delta V_{\mathrm{gap,BG}}$. The dashed line is a linear fit to the data points of the five wider constrictions. The inset shows all the data points, with the linear fit extrapolated towards larger gap sizes.}
\end{figure}

Fig. \ref{fig:Fig3}(a) shows the transport gap in back gate voltage $\Delta V_{\mathrm{gap,BG}}$, determined as shown in Fig. \ref{fig:Fig2}(b) and described in section \ref{sec:tg}, as a function of the constriction width. The error bars in horizontal direction result from the resolution of the SFM scans, while the vertical error bars are determined by applying the procedures to different measurements of the same constriction, and using different ranges for smoothening and fitting. We find a decay of the gap size with increasing constriction width, similar to what has been observed by Han et al. \cite{han_prl2007} for the size of the transport gap in energy. For the longer constrictions with $L=200$~nm and especially $L=500$~nm, $\Delta V_{\mathrm{gap,BG}}$ is larger than a value one would expect for a 100~nm long constriction of the same width. This can be understood by the fact that increasing the width of the constriction or decreasing its length increases the probability of at least one percolating conductive path through the constriction, and therefore decreases the region in energy where transport is governed by localization. 
For the constrictions with $W=75$~nm and $L$=100~nm, $W$=100~nm and $L$=100~nm respectively $W$=85~nm and $L$=200~nm, $\Delta V_{\mathrm{gap,BG}}<0$. The value of $\Delta V_{\mathrm{gap,BG}}$ includes an offset which depends on the conductance value chosen to measure the distance between the fitted lines. With our choice of taking the intersection at $G=0$, a negative value of $\Delta V_{\mathrm{gap,BG}}$ means that the intersection point of the two fitted lines lies at a positive conductance value. In these cases, even though the conductance is reduced due to localized states in the constriction, it is never completely suppressed. 

Fig. \ref{fig:Fig3}(b) displays the size of the energy gap in bias direction $E_{\mathrm{gap}}$, determined from the largest diamond in the gap region, as a function of the gap extension $\Delta V_{\mathrm{gap,BG}}$ in back gate voltage. $E_{\mathrm{gap}}$ increases approximately linearly with $\Delta V_{\mathrm{gap,BG}}$ for the five rather wide constrictions. The dashed line indicates the result of a linear fit, taking into account the data points of those five constrictions and their error bars, given by 
\begin{equation}
E_{\mathrm{gap}}[\mathrm{meV}]=(1.1\pm0.3)~\frac{\mathrm{meV}}{\mathrm{V}} \cdot \Delta V_{\mathrm{BG,gap}}+(4.5\pm0.6)~\mathrm{meV}.
\end{equation}
An extrapolation of the fit to larger gap sizes shows that the gap of the 30 nm wide constriction also agrees well with the fitted line.

Even though the slope of the fit relates the back gate voltage to an energy, it does not represent the lever arm of the back gate on the constriction, as the latter can be determined from the diamond measurement in Fig. \ref{fig:Fig2}(d). The slope can rather be understood as describing the envelope of the diamond-shaped region of suppressed conductance in Fig. \ref{fig:Fig2}(c). The proportionality of $E_{\mathrm{gap}}$ and $\Delta V_{\mathrm{gap,BG}}$ means that this slope does not depend on the constriction geometry.
The connection between  $\Delta V_{\mathrm{gap,BG}}$ and $E_{\mathrm{gap}}$ relates two energy scales characterizing the disorder potential. The quantity $E_{\mathrm{gap}}$ is presumably related to the charging energy of the individual islands. The gap size in back gate voltage $\alpha \Delta V _{\mathrm{gap,BG}}$ is a measure of the energy interval in which localized electrons govern transport at the Fermi energy. It may be related to the average amplitude of the potential fluctuations due to disorder.\cite{martin_naturephys2008} However, the reason for this apparently linear relation between this two quantities is not yet understood in detail.
 
\subsection{Influence of an applied side gate voltage}\label{sec:sg}

\begin{figure}
\includegraphics[width=0.45\textwidth]{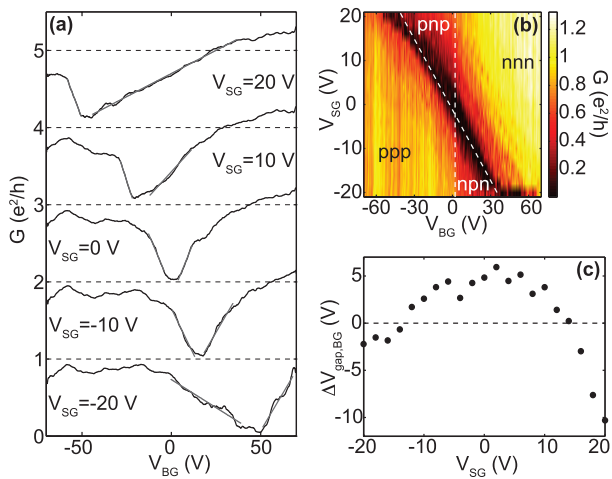}
\caption{\label{fig:Fig4} Constriction with dimensions $W=65$~nm, $L=100$~nm (a) Conductance as a function of $V_{\mathrm{BG}}$ for different fixed side gate voltages $V_{\mathrm{SG}}$. The traces are smoothened over 1.5 V in $V_{\mathrm{BG}}$, and are offset by $e^{2}/h$. The grey lines indicate the linear fits used to determine $\Delta V_{\mathrm{gap,BG}}$. (b) Color plot of the conductance as a function of $V_{\mathrm{BG}}$ and $V_{\mathrm{SG}}$. The vertical white dashed line indicates the transition from electrons to holes in the leads, while the diagonal line separates regions with different carrier type in the constriction. (c) $\Delta V_{\mathrm{gap,BG}}$ as a function of applied $V_{\mathrm{SG}}$. The dashed line indicates $\Delta V_{\mathrm{gap,BG}}=0~\mathrm{V}$.}
\end{figure}

The effect of a voltage applied to both side gates is studied for the constriction with $W=65$~nm and $L=100$~nm.
Fig. \ref{fig:Fig4}(a) displays the smoothened conductance as a function of $V_{\mathrm{BG}}$ for different values of $V_{\mathrm{SG}}$ applied to both side gates. The position of the conductance minimum in $V_{\mathrm{BG}}$ shifts with an applied side gate voltage $V_{\mathrm{SG}}$, according to the relative lever arms on the constriction $\frac{\alpha_{\mathrm{BG}}}{\alpha_{\mathrm{SG}}}=\frac{|\Delta V_{\mathrm{SG}}|}{|\Delta V_{\mathrm{BG}}|}\approx \frac{40}{77}\approx \frac{1}{2}$. The same effect can be observed in the color plot in Fig. \ref{fig:Fig4}(b) displaying the conductance as a function of both $V_{\mathrm{BG}}$ and $V_{\mathrm{SG}}$. 

Another effect visible in Fig. \ref{fig:Fig4}(a) is the asymmetry in the slopes around the gap for high enough side gate voltages. Further insight can be obtained by having a closer look at Fig. \ref{fig:Fig4}(b). The vertical dashed line represents the position of the charge neutrality point in the graphene leads, assuming that the applied side gate voltage has no significant influence on the leads. The diagonal dashed line indicates the charge neutrality point in the constriction. On one side of the transport gap, the current is reduced due to the higher resistance in the leads. This leads to the asymmetry observed in Fig. \ref{fig:Fig4}(a).\cite{huard_prl2007,stander_condmat2007,williams_science2007,oezyilmaz_prl2007,gorbachev_nanolett2008}

Fig. \ref{fig:Fig4}(c) shows the width of the transport gap in back gate voltage, as defined in section \ref{sec:tg}, as a function of applied $V_{\mathrm{SG}}$. The gap is clearly defined around $V_{\mathrm{SG}}=0$~V, but its width decreases with increasing $|V_{\mathrm{SG}}|$ until even $\Delta V_{\mathrm{gap,BG}}<0$ for $|V_{\mathrm{SG}}|>15$~V. For such high side gate voltages, due to screening there is a strong gradient in potential across the width of the constriction. Therefore, it is not possible for the Fermi level to be in the vicinity of the charge neutrality point in the whole constriction at the same time. Localization of charge carriers only happens in a certain region of the constriction, but never over the whole width. This localization still leads to a reduction of the conductance, but transport is no longer completely suppressed.

\subsection{Magnetic field effect}\label{sec:B}

\begin{figure}
\includegraphics[width=0.45\textwidth]{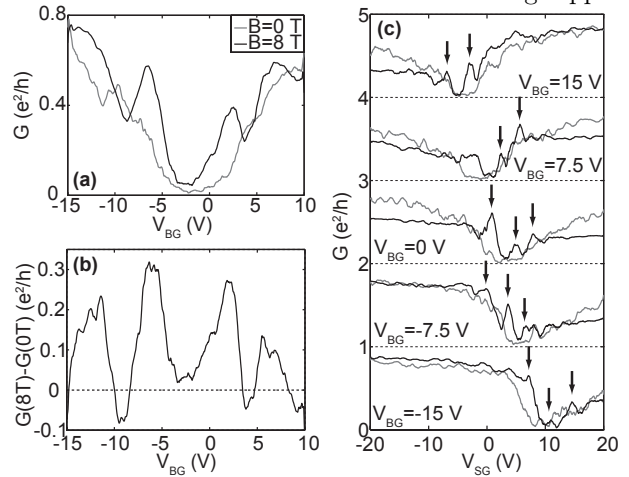}
\caption{\label{fig:Fig5} Constriction with dimensions $W=65$~nm, $L=100$~nm (a) Conductance as a function of $V_{\mathrm{BG}}$, measured at $B=0$~T (grey trace) and $B=8$~T (black trace). Both traces are smoothened over 2~V in $V_{\mathrm{BG}}$. The side gates are grounded. (b) Difference between the smoothened conductanced for 0 and 8~T. The dashed line indicates $G(0~\mathrm{T})-G(8~\mathrm{T})=0$. (c) Conductance as a function of $V_{\mathrm{SG}}$ for different $V_{\mathrm{BG}}$. The traces are smoothened over 1~V in $V_{\mathrm{SG}}$ and offset by $e^{2}/h$.}
\end{figure}

The same constriction ($W=65$~nm and $L=100$~nm) is used to study the effect of an external magnetic field applied perpendicular to the graphene plane. Fig. \ref{fig:Fig5}(a) shows the smoothened conductance as a function of $V_{\mathrm{BG}}$, measured at $B=0$~T (grey trace) and $B=8$~T (black trace). The charge neutrality point is shifted to $\approx-2$~V due to a change of the sample with time. There are some regions where the conductance is significantly increased  by the magnetic field. This effect is studied in detail in Fig. \ref{fig:Fig5}(b) displaying $G(B=8~\mathrm{T})-G(B=0~\mathrm{T})$. Peaks with a spacing in $V_{\mathrm{BG}}$ of about 7~V can be observed. They could be explained by the existence of Landau levels at high magnetic fields. For distinct back gate voltages, the current is expected to flow in edge channels with strongly reduced backscattering, therefore increasing the conductance. This is in agreement with previous quantum Hall effect measurements on a Hall bar for the same magnetic field.\cite{molitor_prb2007} The gate voltage region of suppressed conductance seems to be reduced for finite magnetic field in comparison to zero field. This effect is the less pronounced the narrower the constriction is. It is unclear whether this effect is related to the nature of charge transport through the constriction at finite magnetic field or to a real change of the gap/potential landscape as a function of B-field.

The edge states could form either only in the leads or in the whole structure, including the constriction. This question is studied in Fig. \ref{fig:Fig5}(c) showing the smoothened conductance as a function of side gate voltage for different $V_{\mathrm{BG}}$, again for $B=0$~T (grey traces) and $B=8$~T (black traces). In the traces for $B=8$~T, similar peaks to the ones in Fig. \ref{fig:Fig5}(a) can be observed. The spacing of these peaks in side gate voltage is $\approx3.5$~V. By taking into account the relative lever arms of back and side gates on the constriction determined in section \ref{sec:sg}, we find that the peaks in the back gate trace and the ones in the side gate traces have the same energy spacing, and therefore probably originate from the formation of edge channels in the constriction itself, and not only in the leads. 

\section{Conclusion}

In conclusion, we have presented measurements on graphene nanoribbons of different widths and lengths. We establish a way to describe the transport gap by its width in back gate voltage, and compare it to the usual measurements of the conductance as a function of bias and back gate voltages for different constriction geometries. This method is used to study the effect of an applied side gate voltage on the transport gap formed in a constriction. We find that in the case of a voltage applied to both side gates, there is still a region of reduced conductance, but the conductance is no longer completely suppressed. In the presence of a high enough magnetic field, we observe the formation of Landau levels in the constriction.


\begin{acknowledgments}
The authors wish to thank M. Hilke and T. Heinzel for helpful discussions. Support by the ETH FIRST lab and financial support from the Swiss Science Foundation (Schweizerischer Nationalfonds, NCCR Nanoscience) are gratefully acknowledged. 
\end{acknowledgments}

\end{document}